\documentclass[ceqn, 10pt]{olplainarticle}

\author[1]{Sam Alexander Martino}
\author[1]{João Morado}
\author[1]{Chenghao Li}
\author[1]{Zhenghao Lu}
\author[1*]{Edina Rosta}
\affil[1]{Department of Physics and Astronomy, University College London}
\affil[*]{Correspondence: e.rosta@ucl.ac.uk}

\usepackage{amsmath}
\usepackage{soul}
\usepackage{textcomp}
\usepackage{booktabs}
\usepackage{csquotes}
\usepackage[sort&compress,numbers,super,square]{natbib}
\usepackage{url}
\usepackage{graphicx,
	        caption,
    	    float}
\graphicspath{ {./images/}
		     }
\newcommand{\newpara}
	{
	\vskip 0.4cm
	}

\keywords{Graph Neural Networks, Markov chains, Molecular Dynamics, Kemeny Constant}

\title{Kemeny Constant-Based Optimization of Network Clustering Using Graph Neural Networks}

\usepackage[pagewise]{lineno}
\usepackage{xcolor}
\usepackage{soul}

\begin{document} 
% \linenumbers
\begin{abstract}
The recent trend in using network and graph structures to represent a variety of different data types has renewed interest in the graph partitioning (GP) problem. 
This interest stems from the need for general methods that can both efficiently identify network communities and reduce the dimensionality of large graphs while satisfying various application-specific criteria.
Traditional clustering algorithms often struggle to capture the complex relationships within graphs and generalize to arbitrary clustering criteria. 
The emergence of graph neural networks (GNNs) as a powerful framework for learning representations of graph data provides new approaches to solving the problem.
Previous work has shown GNNs to be capable of proposing partitionings using a variety of criteria. However, these approaches have not yet been extended to Markov chains or kinetic networks. 
These arise frequently in the study of molecular systems and are of particular interest to the biomolecular modeling community.
In this work, we propose several GNN-based architectures to tackle the GP problem for Markov Chains described as kinetic networks. 
This approach aims to maximize the Kemeny constant, which is a variational quantity and it represents the sum of time scales of the system.
We propose using an encoder-decoder architecture and show how simple GraphSAGE-based GNNs with linear layers can outperform much larger and more expressive attention-based models in this context.
As a proof of concept, we first demonstrate the method's ability to cluster randomly-connected graphs. We also use a linear chain architecture corresponding to a 1D free energy profile as our kinetic network.
Subsequently, we demonstrate the effectiveness of our method through experiments on a data set derived from molecular dynamics. We compare the performance of our method to other partitioning techniques such as PCCA+.
We explore the importance of feature and hyperparameter selection and propose a general strategy for large-scale parallel training of GNNs for discovering optimal graph partitionings.
\end{abstract}
\maketitle         

\section{Introduction}
Markov chains (MCs), and their extensions, have become a key tool across multiple domains to model dynamical systems\cite{NatureMCCommunityDetection, 5GreatestMCs}. In their simplest form, MCs incorporate sequential data as a Markovian process into a stochastic model, where the probability of observing the next event depends solely on the preceding event. By eliminating the long timescale dependencies that often occur in complex systems, MCs offer a robust and formal method to interpret high-dimensional data sets in the emergent era of big data. The simplicity, transferability, and capability of MCs reduce the difficulties associated with modelling and interpreting ordered dynamics, and have led to a wide range of use cases. These applications encompass modeling character sequences in computer science and genomics\cite{plagMC, WordsMarkov}, predicting web-user activity\cite{insidePagerank}, and modelling time-dependent processes in physics and chemistry\cite{ChemicalApplications, MSMsPande}, to name just a few.
\newpara
An application that has widely adopted these techniques is biomolecular simulation, where the Markov state model (MSM) formalism has emerged as a strategy for interpreting the results of molecular dynamics (MD) simulations as transitions between aggregated conformational states\cite{MSMsProteinFolding, PathSamplingPande, pan2008building}. The surge in popularity of MSMs mirrors the increased ability to quickly generate large quantities of MD data, providing a useful tool to condense massive data sets into a small and intuitive probabilistic format. MSMs offer intuitive descriptions of complex dynamic systems as transitions between metastable minima and saddle points serving as transition states\cite{LindaPaper}, helping to elucidate the underlying functional kinetics of interest\cite{MSMGibbs, bowman2015discovery}. Over the past few decades, several domain-specific advances led to the development of techniques to automatically generate MSM representations of molecular systems. Simultaneously, these techniques have been employed to further sample the multitude of different configurations of biomolecular systems\cite{mardt2018vampnets, LeveragingMSMs}.
\newpara
Despite being smaller than the huge data sets from which they originate, many MSMs remain high-dimensional, potentially consisting of hundreds of states. This large number of states is often unmanageable for downstream tasks such as drug design and biomolecular engineering\cite{art2science}. As size often becomes the limiting factor to establishing an intuitive understanding of systems, many different approaches have been proposed to reduce the size of MCs and generate optimal ones \cite{deuflhard2005robust, hummer2015optimal, GerhardFolding, GPPCA}. However, the ability to generate large data sets is only increasing, as modern coarse-graining protocols\cite{Greg} and new developments in machine learning\cite{Cecil} allow for larger time scales to be probed.
Traditional graph partitioning (GP) algorithms, such as spectral clustering, k-means, or multilevel methods, and their derivatives, have been widely employed to address this problem \cite{li2016effect, nagel2023toward, roblitz2013fuzzy}. However, these algorithms often face limitations in capturing the complex and intricate dependencies on global structure in MCs. Additionally, the generality of these methods means that they often disregard the critical dynamic information held within MCs. Alternative Markov-Chain Monte Carlo (MCMC) approaches have been applied for dimensionality reduction on MCs, which can be used with most clustering criteria \cite{MCMC1, MCMC2, MCMC3}. Nevertheless, these approaches encounter difficulties in efficiently sampling the most relevant regions of partitioning space, resulting in impractical computation times and frequent overlooking of the most optimal solutions.
\newpara
In this paper, we introduce machine learning (ML) models based on graph neural networks (GNN) to optimize dynamical network clusterings. 
We build upon previous work \cite{VladPaper}  for reducing the dimensionality of MCs, and propose ML architectures leveraging recent developments in using GNNs to solve the more general GP problem\cite{shaham2018spectralnet, kawamoto2018mean, liao2018graph}. Although several existing approaches use GNNs to minimize partitioning cuts\cite{gap} or modularity\cite{DMoN}, these applications do not extend to kinetic graphs or optimization based on dynamical criteria. Additionally, these approaches mostly focus on providing approximate partitionings for large graphs rather than exploring their potential for identifying optimal partitionings in smaller networks. We extend these methods by representing MCs as kinetic networks and show how performing gradient descent on a large number of randomly initialized small GNNs can yield optimal network partitionings using elaborate optimization criteria. As a metric to quantitatively assess the degree to which a clustering preserves the underlying dynamics of the MSM, we use the change of Kemeny constant (KC). KC is an inherently dynamical quantity derived from  mean first passage times (MFPTs) associated with any kinetic network, including those used within biomolecular simulations\cite{MSMs4MD}. We also compare our clustering results with a commonly used kinetic network clustering method for MSMs, the robust Perron-cluster analysis (PCCA+), which does not use the KC as the optimization criterion\cite{kube2007coarse}.

The proposed MC clustering method can produce a faithful low-dimensional replication of a system's underlying dynamics, while also providing a quantifiable measurement related to the amount of kinetic information lost. Along with outlining a general strategy for the large-scale training of GNNs for graph partitioning, we show our method's performance when minimizing the change in KC on several different kinds of synthetic graphs and an example derived from MD simulations. We also test more complex GNN architectures, showing how large-scale training of simple GNNs is the most effective for the task and hypothesizing why this is the case.

\section{Theory}
\label{sec:theory}

\subsection{Graphs}
A graph ${G}$ with $N$ nodes can be described as a set of vertices ${V} = \{v_1, ..., v_{N}\}$ and edges $e_{ij} \in {E}$, which are often represented as a binary adjacency matrix, $\mathbf{A}$, where:

\begin{equation}
    A_{ij} = 
        \begin{cases}
        1, & \text{if}\ e_{ij} \in E \\
        0, & \text{otherwise}
    \end{cases}
\label{eq:adj_matrix}
\end{equation}
Each node, $v_i$, can be associated with a $d_{\text{feat}}$ dimensional feature vector, $F_i$, similarly admitting a matrix representation as a $(N \times d_{\text{feat}})$ matrix, $\mathbf{F}$. These features can be derived from the underlying data or generated at run time in the case of a featureless graph.
This makes it possible to define a general graph structure as:
\begin{equation}
\label{graph_structure}
{G} = ({V}, {E}, {F})    
\end{equation}
Therefore, in practice, a complete specification of the graph structure can be achieved solely using the matrices $\mathbf{A}$ and $\mathbf{F}$.
\newpara

In what follows, the terms clustering and partitioning are used interchangeably to refer to the GP problem. The general partitioning problem consists of dividing the nodes of $G$ into $M$ discrete non-overlapping subsets, $L \in \{1, \ldots,M\}$, such that specific criteria are met. These objectives can be general and applicable to nearly all graphs, and include, \emph{e.g.}, optimizing connectivity, minimizing inter-cluster edges, or maximizing intra-cluster similarities. Furthermore, these objectives can also be domain-specific, based on the underlying model the network represents, such as how the KC represents the dynamical information of the system.
The $(N \times M)$ partitioning assignment matrix, $\mathbf{S}$,  assigns the nodes of $G$ into $M$ distinct states, such that:

\begin{equation}
    S_{iL} = 
    \begin{cases}
        1, & \text{if}\ {v}_i \in L \\
        0, & \text{if}\ {v}_i \notin L
    \end{cases}
\label{eq:assign_matrix}
\end{equation}
Since the discrete nature of the assignment matrix poses difficulties for gradient-based methods, it is common to relax this constraint and work with continuous probabilities instead. In practice, this means introducing a soft assignment matrix instead, where $S_{iL}$ now denotes the probability of node $v_i$ belonging to cluster $L$, subject to the conservation of probability:
$$ \sum^M_{L} S_{iL} = 1 \ \forall \ v_i \in V $$
where $L$ runs over all $M$ possible clusters.

Some of the most commonly used clustering criteria used for the GP problem are the modularity, minimum cut, and Davies-Bouldin index (DBI), which are defined as follows: \\

\textbf{Modularity}: The modularity, $Q$, measures how the connectivity of a graph partitioning differs from what would be expected in a purely random graph with the same node degree distribution. Consequently, this metric provides a quantitative assessment of how statistically surprising a proposed partitioning is, with higher values implying the presence of underlying communities within the data \cite{newman2006modularity}. The modularity is calculated using the equation:
\begin{equation}
\label{modularity}
    Q = \frac{1}{2m} \sum_{ij}^N \sum_{L}^M \bigg[ A_{ij} - \frac{d_i d_j}{2m} \bigg] \ {S}_{iL}{S}_{jL}
\end{equation}

where $m$ is the total number of edges in the graph, $d_i$ and $d_j$ are the degrees of nodes $v_i$ and $v_j$, respectively, $A_{ij}$ is given by Eq. (\ref{eq:adj_matrix}), and $S_{iL}$ and $S_{jL}$ are given by Eq. (\ref{eq:assign_matrix}). Networks characterized by high modularity exhibit dense intra-cluster node connections, but sparse inter-cluster node connections. While there have been criticisms of the underlying assumptions for modularity, it is a widely used and useful partitioning criterion in many areas\cite{critMod, Mod2, Mod3}.

\textbf{Minimum Cut}: The minimum cut, $C$, of a given partitioning is the number of inter-cluster edges, which can be calculated using:
\begin{equation}
% Ugly equation
    \label{mincut}
    C = \sum^N_{i}\sum^N_{j \neq i} \sum^M_{K} \sum^M_{L\neq K} S_{iK} S_{jL} A_{ij}
\end{equation}

where all indices and matrix values are given as previously defined. The simplicity of this criterion has led to its popularity across multiple domains, spurred on by its proved equivalence to the max-flow criterion \cite{MaxFlowMinCut}. It stands as one of the most studied partitioning criteria \cite{MinCut1, MinCut2, MinCut3}, alongside its further generalisations commonly used to obtain more balanced clusters with minimal cuts \cite{balancedcut, bc2}.
Intuitively, the minimum cut partitioning of a network corresponds to the partitioning with the least amount of inter-cluster edges.

\textbf{Davies-Bouldin index}: The DBI quantifies the average similarity between clusters\cite{DBMetric}, calculated based on the distance between clusters relative to the size of the clusters themselves, such that the DBI formula reads:

\begin{equation}
\text{DBI}=\frac{1}{M} \sum_{L}^M \max _{K \neq L}  \left[ \frac{\sigma_K+\sigma_L}{\Delta_{KL}} \right]
\end{equation}

where $L$ runs over all possible $M$ clusters, $\Delta_{KL}$ denotes the distance between cluster centroids, \emph{i.e.} $\Delta_{KL}=\left\lVert  \mathbf{B}_K-\mathbf{B}_L  \right\rVert$, and $\sigma_L$ represents the cluster diameter, calculated as the average distance between the feature vectors and the centroid of the respective cluster:

\begin{equation}
\sigma_L = \frac{1}{N} \sum_i^N \left\lVert  \mathbf{F}_i-\mathbf{B}_L  \right\rVert
\end{equation}

where $\mathbf{F}_i$ represents the feature vector of node $i$, and the centroid $\mathbf{B}_L$ is computed by averaging the features vectors of all $N$ nodes $i$ within cluster $L$. A lower DBI value (bounded from below by 0) indicates better partitioning, as it suggests higher inter-cluster dissimilarity in comparison to intra-cluster similarity. We employed the DBI score as a metric for  assessing clustering performance.

\subsection{Markov Chains}

The dynamical clustering criterion based on KC we use in this paper is derived from the kinetic properties of an MC.
In previous work, we explored the relationship between the MFPTs and KC, providing thorough derivations of all the quantities presented below. Here we briefly summarise the key results required and direct the reader to our previous paper for a detailed reference\cite{AdamPaper}.

In a discrete-time, homogeneous MC comprising of $n$ states, the probability vector of a random walk occupying any of the $n$-states at time $t$, $\mathbf{p}(t)$, can be expressed using a stochastic rate matrix, $\mathbf{K}$, and a set of initial occupation probabilities, $\mathbf{p}(0)$, such that:

\begin{equation}
\mathbf{p}(t) = e^{\mathbf{K}t}  \mathbf{p}(0)    
\end{equation}

where $t=l \tau$, since in the discrete case transitions between states occur at integer multiples of a given time interval $\tau$. 
The quantity $\mathbf{Q}(\tau)=e^{\mathbf{K}\tau}$ is known as the Markov matrix and is commonly used as the fundamental representation of a Markov chain. When using this representation, time propagation corresponds simply to repeatedly multiplying  $\mathbf{Q}$ by the current state of the system. 
To ensure the conservation of probability, we have:

\begin{equation}
\sum_j K_{ji} = 0,   \forall i    
\end{equation}

The matrix $\mathbf{K}$, or similarly $\mathbf{Q}$, can be calculated from experimental or simulated transition data, such as MD trajectories, or a series of string-like sequences typically used in genomics. 
$\mathbf{K}$ fully defines the dynamic properties of the MC and can be used to derive several useful quantities relating to its underlying dynamics. Assuming detailed balance, the spectral decomposition of $\mathbf{K}$ into left, $\psi^l_i$, and right, $\phi^l_i$, eigenvectors and corresponding eigenvalues, $\lambda_l$, is given by:

\begin{equation}
\mathbf{K} = \sum^N_{l=1} \lambda_l \psi^{(l)} \phi^{(l)}
\end{equation}

where the index $l$ orders the eigenvalues in descending order. These eigenvalues are upper-bounded by 0 and are guaranteed to be real due to the detailed balanced assumption.
The eigenvector corresponding to the largest right eigenvector, $\psi^{(1)}$, is known as the equilibrium probability and is commonly denoted by $\mathbf{p}^{\text{eq}}$, corresponding to the distribution the system converges to as $t \to \infty$.
The other eigenvectors contain useful kinetic information corresponding to the relaxation and mixing times of the MC.
These eigenvectors can be used to derive an expression for the average time it takes for a random walk starting from state $i$ to first arrive at state $j$, \emph{i.e.}, the MFPTs of an MC, denoted $t_{ji}$:

\begin{equation}
t_{ji} = \frac{1}{p_j^{\text{eq}}} \bigg[
                            \sum_{l>1} \frac{ 1 }{|\lambda_l|} 
							    \psi^{(l)}_j( \phi^{(l)}_j - \phi^{(l)}_i)
				   \bigg]
\end{equation}

which can be alternatively expressed in matrix form as:
\begin{equation}
t_{ji} = \frac{1}{p_j^{\text{eq}}} 
\bigg[
        (\mathbf{p}^{\text{eq}} \mathbf{1}^T_N - \mathbf{K})^{-1}_{jj} 
        - (\mathbf{p}^{\text{eq}} \mathbf{1}^T_N - \mathbf{K})^{-1}_{ji}
\bigg]
\end{equation}

KC is the surprising result that the sum over all states of the product of the MFPTs with the equilibrium probabilities, $p^{\text{eq}}_j$, is a constant value for all $i$, \emph{i.e.}:

\begin{align}
    \zeta = \sum_j p^{\text{eq}}_j t_{ji} = \sum_j \sum_{l > 1} \frac{1}{|\lambda_l|} \psi^{(l)}_j ( \phi^{(l)}_j - \phi^{(l)}_i) = \sum_{l > 1} \frac{1}{|\lambda_l|}
\end{align}

where the fact that  $\sum_j \phi^{(l)}_j$ $\psi^{(l)}_j=1, \forall l$ and $\sum_j \psi^{(l)}_j=\delta_{l1}$ was used \cite{AdamPaper}.
Intuitively, therefore, the Kemeny constant is an important quantity to measure the overall timescales present in a dynamical system, as it is equal to the sum of all relaxation times, $1/|\lambda_l|$. 

% This constant provides an inherently dynamical scalar quantity, ideal for gradient-based optimisation methods. % Removed this to not repeat below
By considering a clustering with the assignment matrix, $\mathbf{S}$, from Eq. (\ref{eq:assign_matrix}), we can define the equilibrium probability of cluster $L$,  $P^{\text{eq}}_L$, as simply the sum of the populations of the individual nodes:

\begin{equation}
P^{\text{eq}}_L = \sum_{i\in L} p_i S_{iL}    
\end{equation}

KC can then be redefined in terms of a partitioned MC's KC, $\zeta_{\textbf{S}}=  \sum_L P^{\text{eq}}_L \hat{t}_{LK}$, where $\hat{t}_{LK}$ is the coarse-grained mean first passage time from state $K$ to $L$. We can define $\hat{t}_{LK}$ via \cite{AdamPaper}:
\begin{equation}
\hat t_{LK}=
\frac{1}{P_K^{\rm eq} P_L^{\rm eq}}\sum_{i \in K, j \in L}p^{\rm eq}_j p^{\rm eq}_i t_{ji}
-\frac{1}{( P_L^{\rm eq})^2}\sum_{i, j \in L}p^{\rm eq}_j p^{\rm eq}_i t_{ji},
\label{eq:MFPT_CG}
\end{equation}
which leads to a variational form compared with the unclustered KC\cite{VladPaper}:
\begin{equation}
	\zeta =  \zeta_{\textbf{S}} + 
	\sum_L^M \frac{1}{P^\text{eq}_L} \sum_{i,j \in L} p^{\text{eq}}_j p^{\text{eq}}_i t_{ji}
	\label{kemeny_cg1}
\end{equation}
Rearranging this equation yields our partitioning metric $\Delta K$:
\begin{equation}
	\zeta - \zeta_{\textbf{S}} = \Delta K = 
	\sum_L^M \frac{1}{P^\text{eq}_L} \sum_{i,j \in L} p^{\text{eq}}_j p^{\text{eq}}_i t_{ji} = 
	\sum_L^M \frac{1}{P^\text{eq}_L} \sum_{i,j } p^{\text{eq}}_j p^{\text{eq}}_i t_{ji} S_{iL} S_{jL}
	\label{kemeny_cg}
\end{equation}
From this equation, it is clear that for the trivial single-state partitioning $M = 1$, $\Delta K=\zeta$ holds, and when $M = N$, the value of $\Delta K$ is 0. 
As KC represents the sum of relaxation times in a system, and a clustering speeds up relaxation times by removing intra-cluster relaxation times, maximizing the KC of a proposed partitioning aims to preserve the original timescales as closely as possible.
From Eq. (\ref{kemeny_cg}) (with the last equality requiring a crisp clustering), we can see that maximizing the KC is identical to minimizing $\Delta K$, and so this provides a useful metric of how kinetically disconnected the clusters are, in a form suitable for gradient-based optimization methods like neural networks.

\subsection{Graph Neural Networks}

To describe GNNs in this work we adopt a message-passing formalism \cite{gilmer2017neural} where a GNN-based encoder maps a node $v_i$ in ${G}$ to a context-aware embedding using $T$ separate layers via the iterative process:
\begin{equation}
    \label{MessagePassing}
    \begin{split}
        m^{t+1}_{i} &= \sum_{j \in \mathcal{N}(v_i)} M_t(h^t_i, h^t_j, e_{ij}) \\
        h^{t+1}_{i} &= U_t(h^t_i, m_i^{t+1})
    \end{split}
\end{equation}

where $t \in \{1, ..., T\}$ denotes the current network depth, $\mathcal{N}(v_i)$ is the local neighbourhood of all nodes connected to node $v_i$ by edges $e_{ij}$, and $h^t_i$ represents the hidden representation of node $v_i$ at depth $t$. For consistency, the initial node feature vectors $F_i$ are represented by $h^0_i$. \\
The functions $M_t$ and $U_t$ are commonly implemented as neural networks, known as the message-passing function and vertex update function, respectively.
Their exact formula depends on the type of graph layer used. \\
Two different graph layers are tested in our methodology: the GraphSAGE\cite{GraphSAGE} and the GATv2\cite{GATv2} layers. The former was chosen to provide a simpler, lower parameter function, which takes fewer resources to train, while the latter provides a more complex layer with the capability to better attune to all nodes. 
GraphSAGE layers use a simple message-passing function consisting of either the sum or mean, and a vertex update function which concatenates these expressions and feeds them into a feed-forward neural network:
\begin{equation}
    m_i^{t+1} = 
    \begin{cases}
        \frac{1}{n} \sum_{j \in \mathcal{N}(v_i)}  h^{t}_j, & \text{if mean} \\
        \underset{j \in \mathcal{N}(v_i)}{\text{max}} \sigma \bigg( \mathbf{W}_{\text{pool}} h^{t}_{j} + b \bigg) & \text{if max}
    \end{cases}
\end{equation}

\begin{equation}
    h_i^{t+1} = \sigma \Bigg( \mathbf{W^k} [ h_i^{t} || m_i^{t+1} ] \Bigg)
\end{equation}
where $[\cdot || \cdot]$ denotes concatenation and $\sigma$ is any non-linear function. \\
GATv2 layers replace this with a formulation based on scaled dot-product attention seen in transformer layers. 
This differs by performing the aggregation step individually for each node in a neighbourhood to calculate a context-aware attention mask, which is then used to perform a weighted vertex update:
\begin{align}
    m({h}^t_i, {h}^t_j) &= \underset{j \in \mathcal{N}(v_i)}{\text{softmax} } \Bigg( \mathbf{a}^T \text{LeakyReLU}(\mathbf{W}\cdot [h^t_i||h^t_j]) \Bigg) \\
    h^{t+1}_i &= \sigma \Big( \sum_{j \in \mathcal{N}(v_i)} m(h^t_i, h^t_j) \cdot \mathbf{W} h^t_j \Big)
\end{align}
which can be generalised across multiple attention heads. \\

\section{Methodology}
\newpara
The proposed method assumes that a given neural network architecture, $\Phi(\theta)$, is sufficiently expressive such that the mappings it explores, from the space of $d_{\text{feat}}$ dimensional node features, $\mathbf{F}\in \mathcal{R}^{N\times d_{\text{feat}}}$, to the space of possible $M$-state partitionings, $\mathbf{S}\in \mathcal{R}^{N \times M}$, includes the optimal solution, $\textbf{S}^{\text{opt}}$, to the GP problem given our clustering criteria, along with a region of parameter space which can be optimized via gradient descent to yield it. Taking inspiration from Monte Carlo approaches to similar problems, we look to determine $\textbf{S}^{\text{opt}}$ by initializing $\Phi$ with several different sets of parameters $\theta$ and optimizing them in parallel using modern ML tools. In this framework, GNNs are an effective Ansatz to begin partitioning graphs using complex criteria. They also provide a generic and intuitive optimization framework with gradient descent. As $\Phi$ is a general model, operating in $\mathbb{R}$ without any problem-specific restrictions on its values or gradients, it can yield the trivial solutions to Eq. (\ref{kemeny_cg}), such as the aforementioned $M=N$ case. To overcome this, we take inspiration from similar work on producing balanced clusters \cite{gap, DMoN}, and added a penalty function to $\Delta K$, which drives the network away from trivial solutions and gives the final form of the loss function, $\mathcal{L}$:

\begin{equation}
    \label{loss_function}
    	\mathcal{L} = 
     \underbrace{ \sum_L^M \frac{1}{P^\text{eq}_L} \sum_{i,j }^N p^{\text{eq}}_j p^{\text{eq}}_i t_{ji} S_{iL} S_{jL}}_{\Delta K}
	 +  \underbrace{ \sum_L^M \frac{\eta}{\sqrt{2 \pi \sigma}} \exp{\left[ -0.5 \left( \frac{\sum_i^N S_{iL} }{\sigma} \right) ^ 2 \right] }}_{\text{Penalty Function} }
\end{equation}

Optimal KC partitionings can have unbalanced subsets, with clusters that contain only a few nodes often being the most interesting transition states\cite{LindaPaper}. Therefore, the form of the penalty function is designed to penalise any invalid clustering (where any cluster has zero nodes), while also allowing the network to converge significantly unbalanced partitions with at least one node.
The final penalty function is based on a Gaussian centered at 0, parameterized in this work using $\sigma = 1.0$ and $\eta = 15$. This means the penalty increases as any cluster approaches 0 nodes and becomes negligible for clusters with 1 or more nodes.

As MCs do not fit into the graph structure outlined in Eq. (\ref{graph_structure}), simple methods were derived for interpreting node connectivity using an adjacency matrix, $\mathbf{A}$, and node feature matrix, $\mathbf{F}$.
The connectivity of the graph was decided using a cutoff method, taking all node pairs $(v_i, v_j)$ with $(Q_{ij}+Q_{ji})/2 > c$, where $c$ is a constant.
In practice, we found $c = 0.015$ to be sufficient for all graphs we worked with, although in other applications this could be changed depending on the density of the resulting network.

\begin{figure}[H]
	\centering
	\includegraphics[width=1\textwidth]{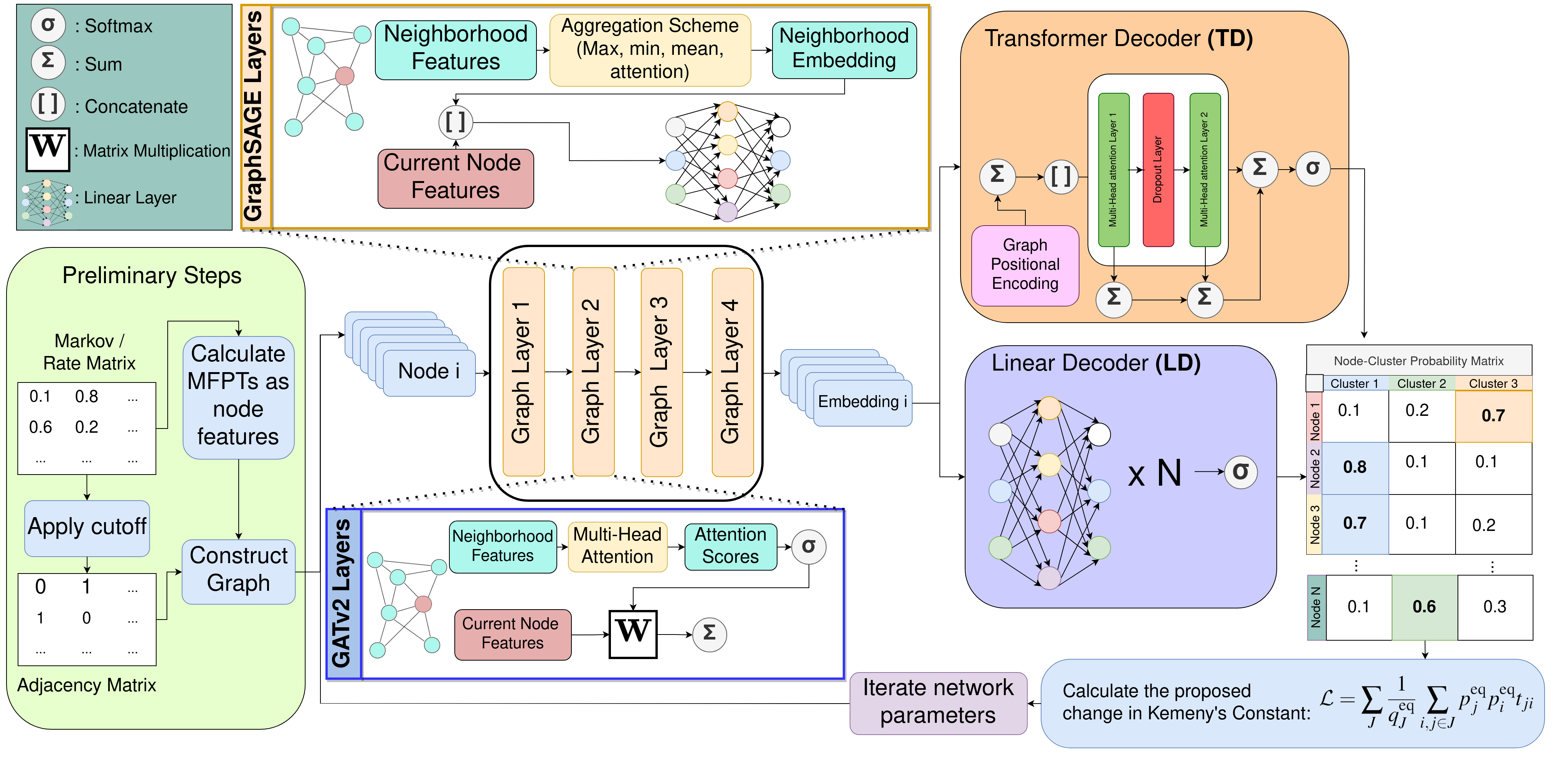}
	\caption{
		Details of all four GNN architectures used, along with a description of the preprocessing required to begin training, together with an overview of the training loop process for a single gradient optimization step. Implementations of the GraphSAGE and GATv2 GNN layers used in the encoder are provided, along with a schematic description of the transformer decoder (TD) and linear decoder (LD) architectures.  Activation functions for the various layers are not included for brevity.
	}
	\label{main-architecture}
\end{figure}

A systematic exploration of various node features was undertaken. 
We tested using the columns of various matrices which can be identified with nodes, such as the adjacency matrix, the rate matrix, the Markov matrix, and the MFPTs, as node features. For each node, we represented its connectivity, transition rates/probabilities to other nodes, or average time required to reach each node, by splitting these matrices into vectors. Alongside these, principal component analysis (PCA), implemented as the singular value decomposition (SVD) of each of these matrices, of all previously mentioned matrices, and the eigenvectors of the rate or Markov matrix were also tested. Finally, we also tested both a simple linear layer, which combined the adjacency, rate/Markov, and MFPT columns for each node, and a trainable embedding layer, similar to those often used as dictionaries in large language models \cite{mikolov2013distributed}. More details regarding the features used are given in SI(A).
\newpara
All neural network architectures tested used a GNN-based encoder, mapping the $d_{\text{feat}}$ dimensional features into an embedding dimension $d_{\text{embed}}$ via a hidden representation of size $d_{\text{hidden}}$. 
Mean pooling was used for the aggregation function in the GraphSAGE layers as it provided better stability during training. 
The encoder was then followed by either a series of linear layers or dot-product attention-based transformer layers of dimension $d_{\text{hidden}}$ as a decoder \cite{DBLP:journals/corr/KlambauerUMH17}. 
When using attention, the output of each transformer layer was normalised and added back to the input before being processed by a feed-forward layer.
Our implementation used two transformer layers with a dropout layer between them. The output of a single trainable graph layer was used as positional encoding before the decoder. % Add the number of heads and such
This transformer-like, attention-based decoder can process a whole clustering at once as a sequence of node embeddings. It was chosen to provide a comparison between a simpler model and one that is larger, more expressive, and better suited to handle highly contextual sequences, such as the node-cluster assignment matrix.
All neural networks used are described in detail in Fig. \ref{main-architecture}.
Building on the simplest model with GraphSAGE layers and the linear decoder (GraphSAGE + LD), we validated this model and extended it in two directions to test whether more complex architectures could improve performance. Besides the GraphSAGE encoder, we also implemented the GATv2 encoder. In addition to the linear decoder, we also implemented a transformer decoder (TD). This results in four models: (i) GraphSAGE layers and the linear decoder (GraphSAGE + LD), (ii) linear decoder with GATv2 layers (GATv2 + LD), (iii) GraphSAGE with transformer decoder (GraphSAGE + TD), and finally (iv) GATv2 with transformer decoder (GraphSAGE + TD).

\begin{figure}[H]
	\centering
	\includegraphics[width=1\textwidth]{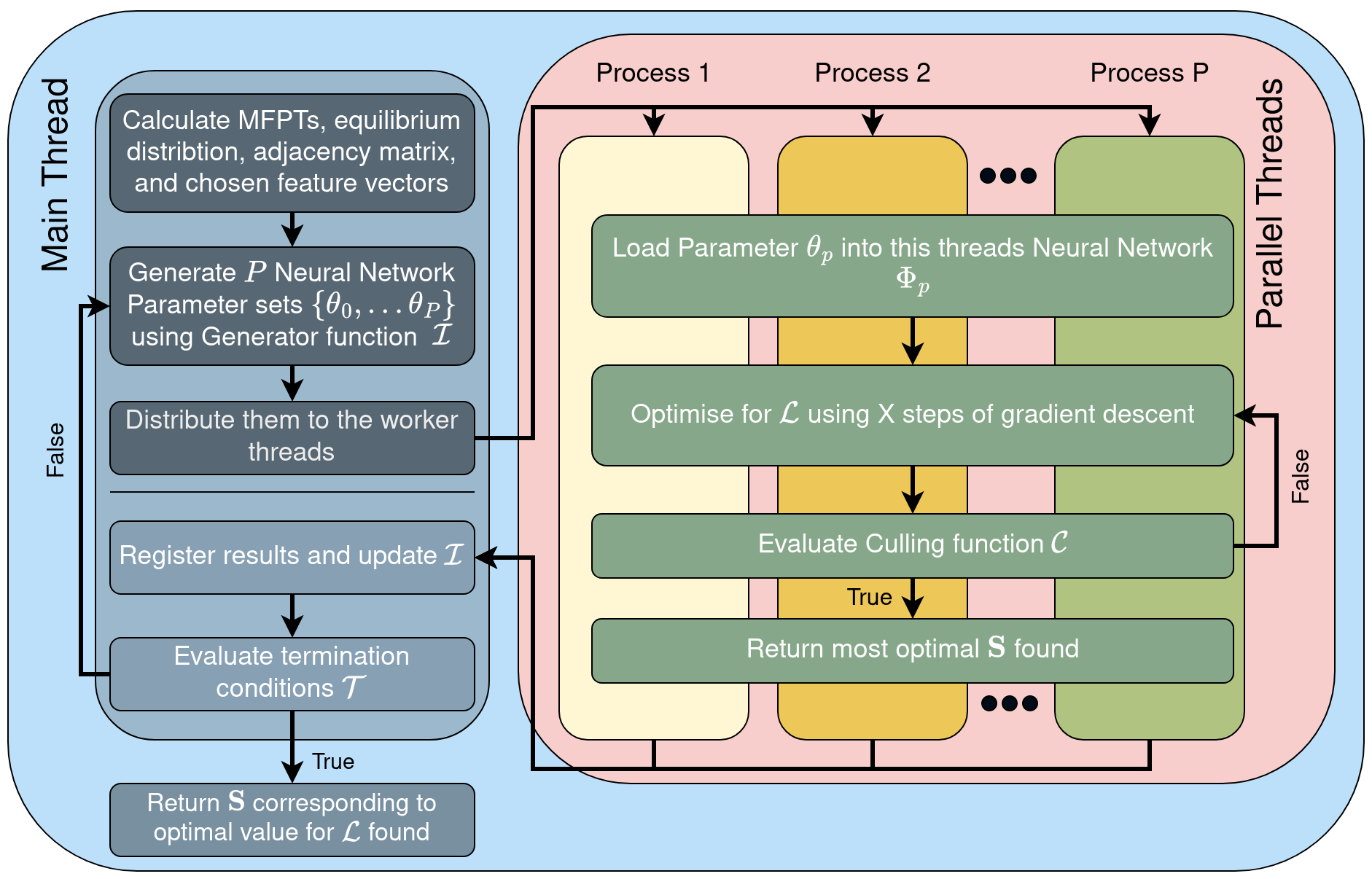}
	\caption{
		Schematic description of the algorithm used to train multiple networks in parallel, and the use of the general functions $\mathcal{I}$, $\mathcal{C}$, and $\mathcal{T}$.
	}
	\label{multiprocess}
\end{figure}

To enable fast and efficient training of the networks, a general parallel methodology for training GNNs for partitioning was established, outlined in Fig. \ref{multiprocess}. 
This requires three additional functions to be defined, as well as a valid loss function and the number of processes $P$: \\
\textbf{Generator Function} $\mathcal{I}$: This function returns valid parameters $\theta$ for a given neural network $\Phi$. This provides the initial starting point in parameter space for optimization, and complex generators can be used to provide a quicker convergence.  In this work, however, simple initialization using both Kaiming and Glorot algorithms was tested. We used the Glorot method due to its slight advantage observed in early tests.\cite{he2015delving, glorot2010understanding} Additionally, we also explored bespoke initializations based on converged models, where a loss function was used that corresponded to the PCCA+ clustering. \\
\textbf{Culling Function} $\mathcal{C}$:  This function determines whether or not a worker thread has finished training a network, and is called after a predetermined number of optimization steps has finished. Observing individual runs, it is common to see neural networks which initialize in local minima of parameter space and are unable to be trained. An effective culling function allows for these types of runs to be terminated early, allowing for more time to be spent in interesting regions of parameter space. To show the validity of the method, all experiments in this paper terminate after a predetermined number of training steps.   \\
\textbf{Termination Function} $\mathcal{T}$:  This function determines when to stop training, and when combined with effective choices for $\mathcal{I}$ and $\mathcal{C}$, can allow for the implementation of complex methodologies for determining the convergence of an experiment. \\
This methodology aims to be as general as possible to make implementing future applications for different criteria straightforward.

\newpara
To test the validity and effectiveness of using GNNs to cluster Markov chains, the method was first tested on synthetic kinetic graphs. 
Firstly, a simple stochastic block model (SBM) \cite{abbe2018community} was employed to generate random networks with 4 neighbourhoods. 
The initial set of networks used sampled inter-cluster connection probabilities from a Gaussian distribution scaled to be in the range $[0, 0.25]$ and set intra-cluster probabilities to a constant 0.5. 
This led to networks with dense clusters, where the optimal partitioning always successfully separated them.
We also designed SBM networks where the optimal clustering was less clear, sampling all connection probabilities from the same Gaussian as before.
For both sets of examples, SBM networks with disconnected nodes were discarded, and the random-walk normalized Laplacian matrix, $\mathbf{L}$, was used to generate rate matrices via:
\begin{equation}
    \mathbf{L} = \mathbf{I} - \mathbf{A}\mathbf{D}^{-1}
\end{equation}
where $\mathbf{I}$ is the identity matrix, $\mathbf{D}$ is a matrix with the degrees of each node along the leading diagonal, and $\mathbf{A}$ is the adjacency matrix given by Eq. (\ref{eq:adj_matrix}).
Results on these synthetic graphs were compared with clusterings and respective $\Delta K$ values obtained from the PCCA+ clustering. We note that as PCCA+ uses a different kinetics-based objective function\cite{VladPaper} that reduced the clustering problem to $M^2$ dimensions, the optimization is readily solvable using Schur decomposition, therefore only a single optimal cluster is obtained using PCCA+.
Each SBM experiment trained 100 independently initialized models for 250 steps of gradient descent on 500 randomly generated graphs, ranging from $N=150$ to $N=300$ nodes in increments of 50.
\newpara
To establish a baseline performance for all models on a single graph, the network was tested on a graph corresponding to a rate matrix derived from an analytical 1D potential simulation\cite{VladPaper}. 
The system consisted of 100 nodes distributed along the potential, which comprised four distinct wells: three with 20 nodes, and one with 40 nodes. The optimal KC partitioning successfully separated all 4 wells into individual partitions. 
This example of a linear chain network architecture was chosen to test the network in situations where connectivity is limited.
All four models were tested 1000 times on this example for 500 gradient descent steps each.
A feature comparison was also performed on this system, where 1000 GraphSAGE + LD models were trained for 500 steps using each set of features described above. \\

Ultimately, the method was tested on a network derived from MD data.
We used the MSM benchmark system of PyEMMA\cite{scherer2015pyemma} corresponding to 500 ns long implicit water MD simulations of a small 5-amino-acid system, here referred to as the pentapeptide system. The simulation initially underwent clustering into 250 states using K-means on tICA projected data, followed by the calculation of a Markov matrix \cite{schwantes2013improvements, perez2013identification}, which served as the input to our method. 
Each experiment consisted of training 1000 models for 500 gradient descent steps to cluster the 250-state MSM into 5 states. Given the relevance of this example to biomolecular simulations, an additional hyper-parameter search was conducted to identify the optimal network architecture. For this purpose, we again trained 1000 GraphSAGE + LD models for 500 steps while varying the number of layers in both the decoder and the encoder, along with the dimensions $d_{\text{hidden}}$ and $d_{\text{embed}}$. Furthermore, we performed a feature comparison following the same methodology as previously described. We also compared the results of this example to the PCCA+ and parallel tempering variational clustering (PTVC), our previous MCMC approach.\cite{VladPaper} Additionally, we used the DBI score with the Markov matrix as the node features vectors as a metric to assess the quality of the partitionings produced.

The neural networks and relevant code were implemented in Python 3.10.6 using PyTorch 2.0\cite{torch} or Tensorflow. \cite{tensorflow2015} 
Network diagrams were generated using the Netgraph library. \cite{NetGraph}
To perform gradient descent steps, the Adam optimizer\cite{kingma2014adam} was employed with a learning rate of $1 \times 10^{-4}$. 
All networks utilized $d_{\text{embed}} = 32$ and $d_{\text{hidden}} = 64$, with 4 GNN layers in the encoder and 2 layers in the decoder, unless otherwise specified.
Due to the numerical sensitivity and dependence of the MFPT calculation on accurate eigenvalues, 64-bit floats were used for all calculations, which were run on the CPU. Each experiment involved the simultaneous execution of 90 processes in parallel, with each process running on a single core. To compute the DBI scores, the scikit-learn implementation was utilized\cite{pedregosa2011scikit}.

\section{Results \& Discussion}
\begin{figure}[H]
	\centering
	\includegraphics[width=1\textwidth]{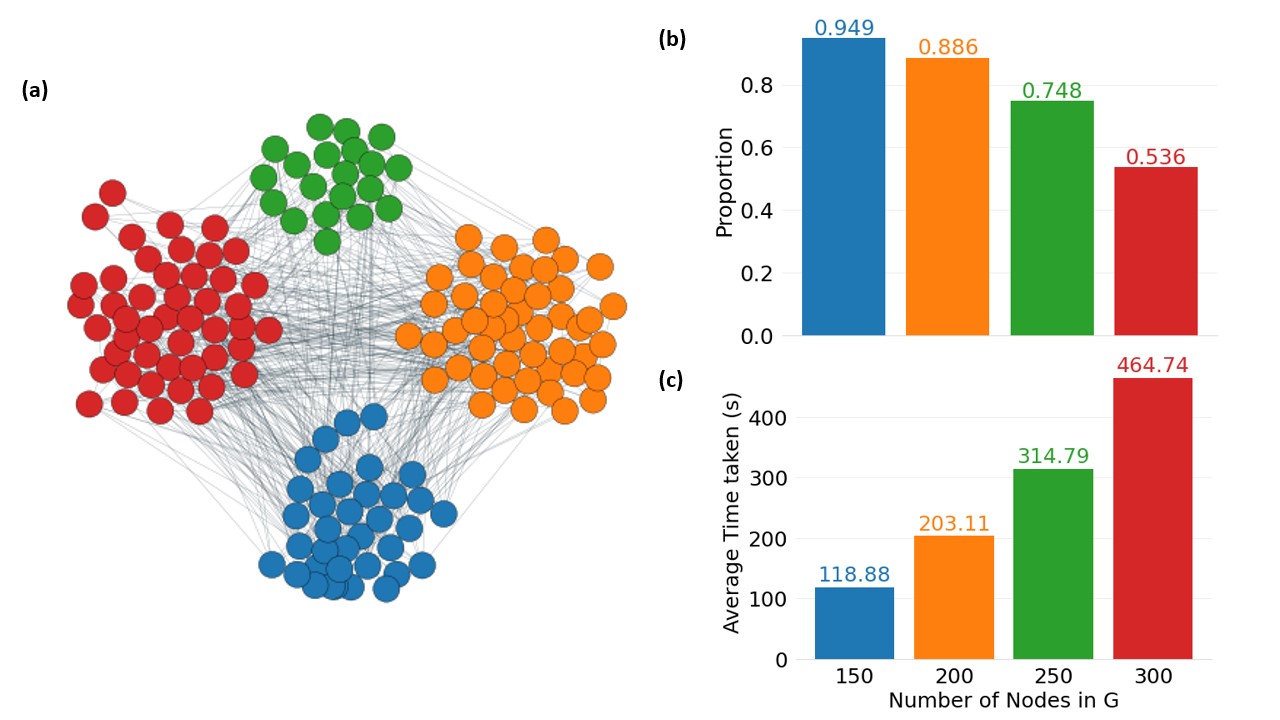}
	\caption{
	The performance of the SBM network clustering for randomly sampled cluster connection probabilities. \textbf{(a)} Example of a partitioning produced by the GNN optimization. \textbf{(b)} and \textbf{(c)} Quantitative assessment of the method's performance over 100 runs for the 500 different networks. \textbf{(b)} The proportion of runs where the method yields a partitioning with a lower $\Delta K$ value than PCCA+. (\textbf{c})  The average time needed in seconds for one training run when limited to run on one 2.20GHz CPU core.
	}
	\label{SBM}
\end{figure}

Our GNN approach successfully found the optimal clustering for all SBM networks in the first set (e.g., Fig. \ref{SBM}a), where the clusters were well-defined, in 100 training runs or less. The optimal cluster matched the PCCA+ solution, confirming the viability of KC as the optimization criterium. Figs. \ref{SBM}b and \ref{SBM}c detail the performance on the second set of SBM graphs featuring a more diverse range of networks. 
Our implementation identified partitionings which better optimized $\Delta K$ than the PCCA+ (Fig. S1). However, the improvement diminished as the number of nodes in the network decreased, as often the clustering became clearer.
The stochastic nature of the GNN-based method however lends itself more to these systems where several viable partitionings have a low $\Delta K$ value. 
In these examples the optimal clustering is close to many other similar solutions, therefore it is helpful to evaluate many initializations to evaluate most of the very low $\Delta K$ partitionings. Violin plot distributions of optimal KC values found are also shown in Fig. S1.

\begin{figure}[H]
	\centering
	\includegraphics[width=1.0\textwidth]{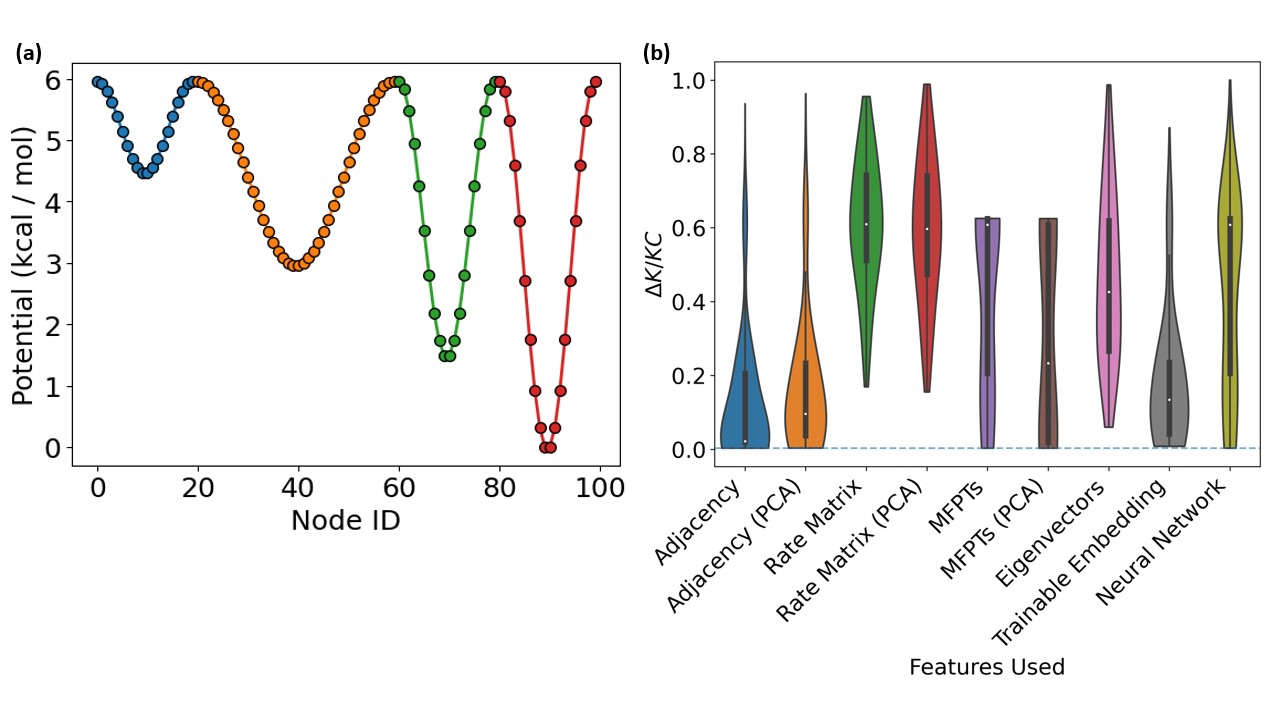}
	\caption{
 (\textbf{a}) Plot showing the 1D potential used for an example of a linear chain network. Nodes are coloured according to the most optimal 4-state clustering for the system. 
 (\textbf{b}) Violin plots for feature comparisons on the 1D potential network. Each experiment was run 1000 times with each set of features. The corresponding distribution of final $\Delta K / KC$ values is shown, along with a dotted blue line signifying the $\Delta K / KC$ value of the optimal solution for each problem. 
	}
	\label{features}
\end{figure}

% \subsection{1D Example}
The 1D potential (Fig. \ref{features}a) was used to test our approach on sparse graphs where connectivity is limited to a linear chain, and the majority of effective $\Delta K$ partitionings lie in a small region of partitioning space. In this regime gradient descent steps can become less effective as the majority of initial partitions are far from the region of interest, and can be unable to escape their local minima. Fortunately, simply training more neural networks can alleviate the problem by providing a wider range of starting points for optimization. 
In future applications, an effective choice of the initialization function $\mathcal{I}$ could speed up convergence, whereby $\mathcal{I}$ can learn and then avoid initializing in uninteresting regions of parameter space.
To further test whether different node features could improve the frequency of successful optimization outcomes, we tested several additional node features, sampling the outcome 1000 times each. In total, we tested nine different node feature types: (i) adjacency matrix, (ii) the PCA vectors created from the adjacency matrix, (iii) rate matrix, (iv) the PCA vectors created from the rate matrix, (v) MFPTs (default in other examples), (vi) the PCA vectors created from the MFPT matrix, (vii) eigenvectors of the rate matrix, (viii) trainable embeddings, (ix) a neural network.
The feature comparison results (Fig. \ref{features}b) show that it is possible to use improved node features, such as the adjacency matrix or a trainable embedding in this case, to achieve a much more successful optimization for the 1D model potential-derived network. \\
When using the rate matrix as node features in this example, the GNN fails to find the correct clustering even once. This may be due to the difficulty in discriminating nodes at high energies (with low populations) based solely on the rate matrix, as these nodes have similar transitions. 
The adjacency matrix, however, provides more information on the node location in the global structure of the network, leading to a noticeable advantage. 
Our GNN implementation allows for simple changes like this to enable defining the node features on a problem-specific basis, therefore giving the method additional viability across a wide range of MCs which vary greatly in structure.\\

\begin{figure}[H]
	\centering
	\includegraphics[width=1\textwidth]{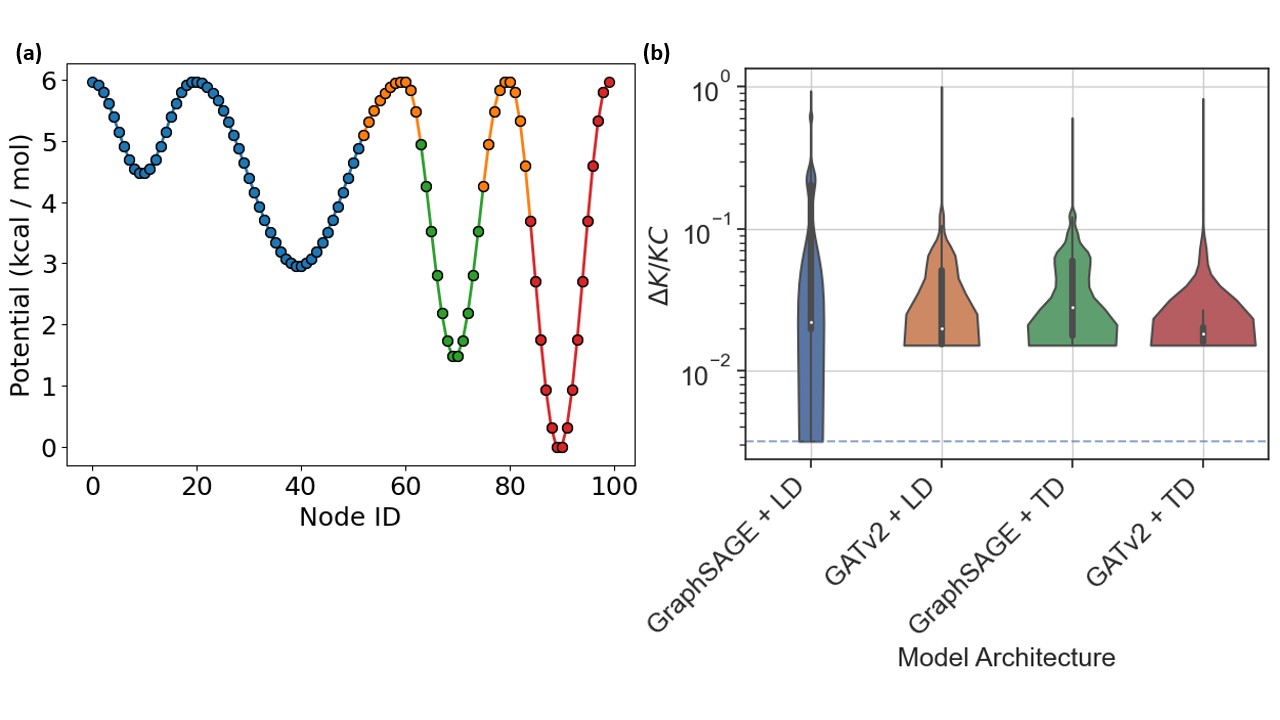}
	\caption{
 \textbf{(a)} The clustering corresponding to the lowest $\Delta K / KC$ values found by all GNN models other than GraphSAGE + LD.
  \textbf{(b)} The distribution of $\Delta K / KC$ values from training 1000 copies of each neural network architecture on the 1D potential derived network. A logarithmic scale is used to highlight the large differences between the different models. The blue dashed line indicates the value associated with the optimal clustering for this system. 
	}
	\label{1D-results}
\end{figure}

We also tested the four GNN architectures on the 1D potential, training 1000 models of each variety using the adjacency matrix as node features. Results are shown in Fig. \ref{1D-results}b and described in Table \ref{table:1D}. We found that simpler models perform better at finding optimal partitionings over more complex models in this framework. Despite the better average performance the GATv2 and TD models provided, no models but the simplest GNN were able to find the optimal clustering over 1000 runs.

The wide variance of the observed results showcases the poor ability of all models to optimize initial partitionings on this network when started from many regions of parameter space, however, in the smaller models this is mitigated by the smaller solution space explored and training time required. 
The large magnitude of some solutions is indicative of a partitioning problem where there is only a small region of interest in partitioning space, leading larger models like the GATv2 and TD models to struggle to find this region, possibly due to the more complex and unclear gradients calculated when the node embeddings are more co-dependent and additional resources and required to train the extra parameters.
There is a sharp cutoff in the performance of all larger models, with the best value attained being consistently around 806 and the corresponding clustering shown in Fig. \ref{1D-results}a. We were unable to pinpoint the exact cause of this lower bound.
We also note that in current MSM building strategies only metastable clusters are identified, unless discretized collective variables are introduced. Typically, a k-means clustering is performed that assigns very rarely sampled transition regions to metastable states. Therefore the optimal clustering even from more complex GNNs (Fig. 5a) is able to separate the three main metastable states, which is in line with current practical MSM applications.
This problem is particularly challenging for GP objective functions that are not derived from kinetics-based measures, and such measures produce qualitatively incorrect optimal clustering, such as Modularity\cite{VladPaper}. Accordingly, the non-optimal solution of the GNNs represented in Fig. \ref{1D-results}a corresponds to a lower DBI value than that of Fig. \ref{features}a, which suggests that the DBI measure is also inadequate for this system (Fig. S2).
The optimal $\Delta K$ clustering, however, is identical to the solution found by PCCA+ in this example and is consistently found by the PTVC method due to the small size of this network.

\begin{table}[H]
\caption{Network architecture comparison for the 1D potential derived network. The four network architectures are compared in terms of the mean (standard deviation) and minimum for 1000 experiments, along with the percentage of times the optimal solution was found. }
\label{table:1D}
\begin{tabular}{|l|c|c|c|}
\hline
\textbf{Model Name} & \multicolumn{1}{l|}{\textbf{Mean}} & \multicolumn{1}{l|}{\textbf{Minimum}} & \multicolumn{1}{l|}{\textbf{Global Minimum Accuracy}} \\ \hline
GraphSAGE  + LD           & 5884.74 ( $\pm$ 8372.62) & \cellcolor[HTML]{FAECCF}\textbf{169.69} & \cellcolor[HTML]{FAECCF}\textbf{5.6\%} \\
GATv2   + LD   & 1988.50 ( $\pm$ 2696.11) & 806.48                                 & 0.0\%                                  \\
GraphSAGE + TD      & 2201.59 ( $\pm$ 1796.45)  & 806.48                                 & 0.0\%                                  \\
GATv2 + TD & \cellcolor[HTML]{FAECCF}\textbf{1508.19  ( $\pm$ 2728.35)}  & 806.48    & 0.0\%                                  \\ \hline
PCCA+ & 169.69  ( $\pm$ 0.0)                                 & 169.69                             & N/A                                  \\ \hline
\end{tabular}

\end{table}
% \subsection{Pentapeptide}

\begin{figure}[H]
	\centering
	\includegraphics[width=1\textwidth]{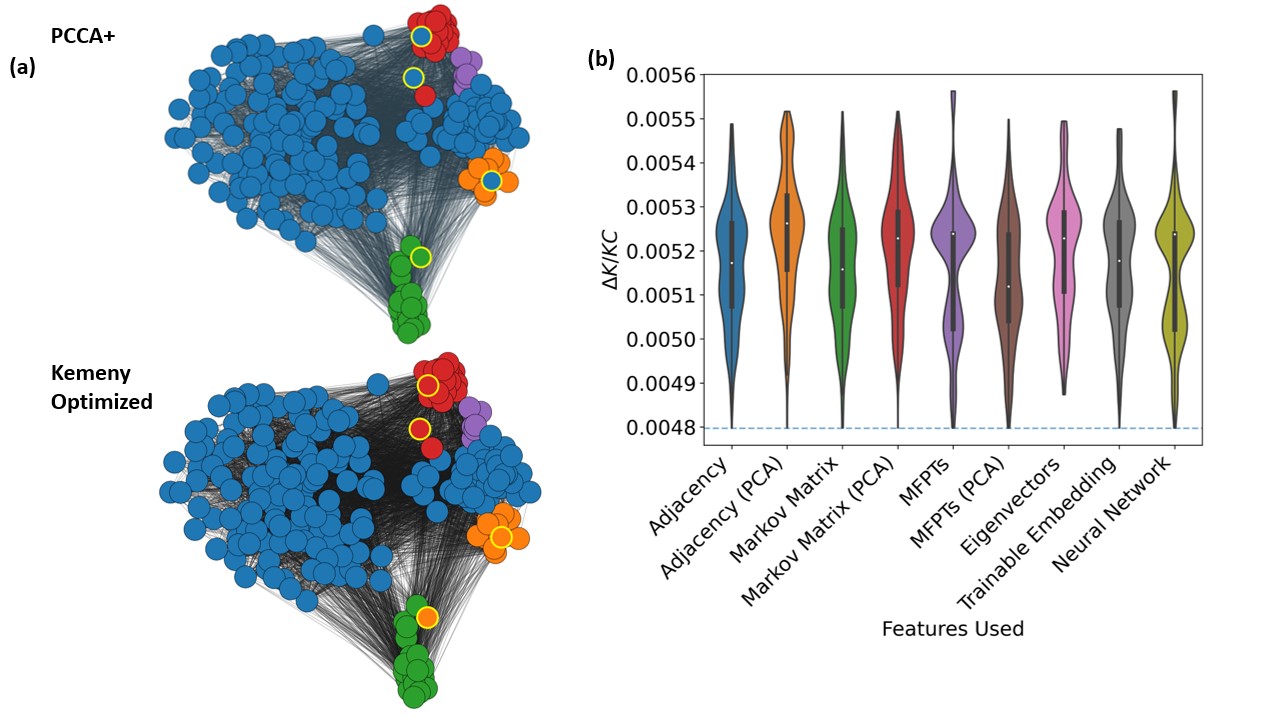}
	\caption{
 \textbf{(a)} Comparison of the PCCA+ solution with the optimal $\Delta K$ clustering for the pentapeptide system, which has DBI scores of 0.823 and 0.758  respectively. Networks are displayed using a force-directed layout weighted with the MFPTs. Nodes which differ between the two clusters are highlighted with a yellow border. \\
  \textbf{(b)} Violin plot distributions for 1000 experiments with each set of nine features. The corresponding distribution of final $\Delta K / KC$ values is shown as a violin plot, along with a blue dashed line for the optimal  $\Delta K / KC$ value. 
	}
	\label{Penta-Diagram}
\end{figure}

\begin{figure}[H]
	\centering
	\includegraphics[width=1\textwidth]{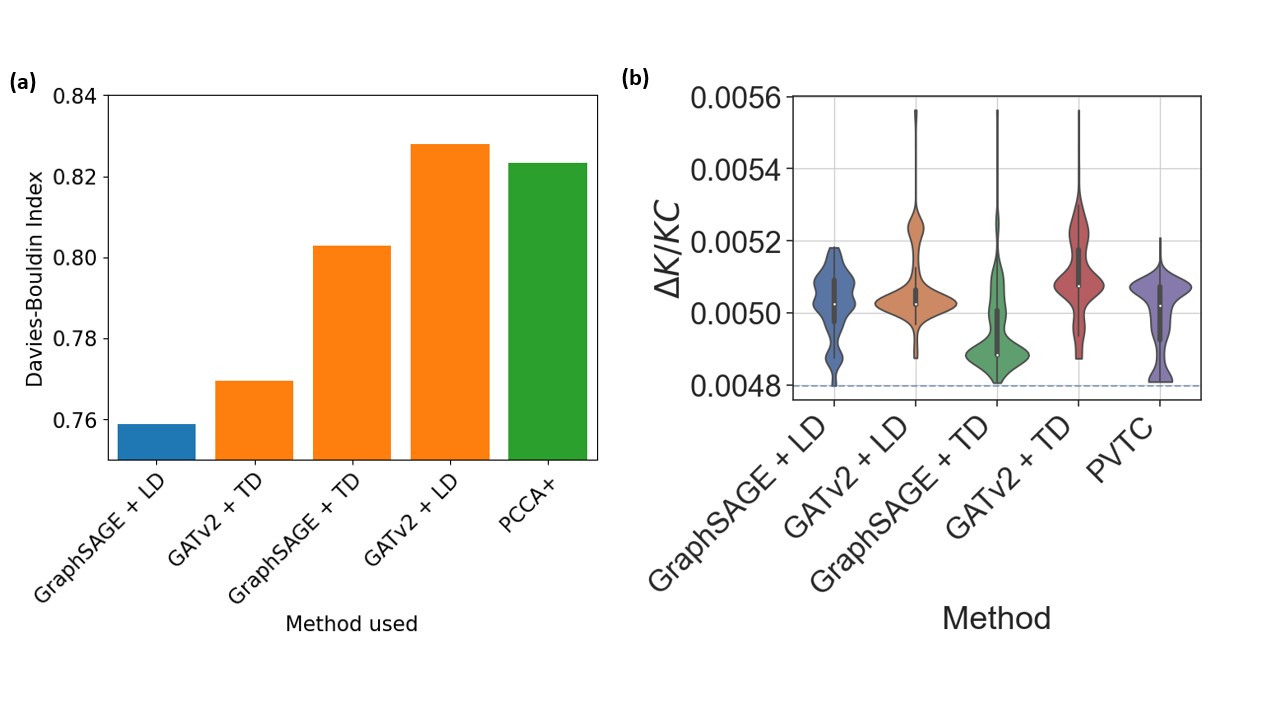}
	\caption{
        \textbf{(a)} A bar-plot showing the DBI values associated with the optimal $\Delta K$ partitionings found by each method used.
		  \textbf{(b)} A violin distribution plot of $\Delta K/K$ values acquired from training 1000 copies of each neural network architecture on the pentapeptide system. The blue dashed line indicates the value associated with the optimal clustering. 
	}
	\label{Penta-Results}
\end{figure}

\newpara
Examples derived from numerical simulations of biomolecular systems often result in noisy derived MCs with dense connections, such as in the pentapeptide case. 
Here, the optimal PCCA+ clustering does not correspond to the optimal  $\Delta K$ solution, with four nodes assigned to different clusters, which are highlighted in Fig. \ref{Penta-Diagram}a and contribute 7.4\% of the overall equilibrium probabilities (Fig. S3).
As shown in Fig. \ref{Penta-Results}b and Table \ref{table:Penta}, again the simplest GraphSAGE and linear decoder method outperforms all other GNN architectures similarly to the 1D example above. 
The PTVC method does not identify the optimal $\Delta K$ clustering in general when using a random initialization, but has been observed to find the correct solution when given the PCCA+ as an initial clustering. 
Here, we also compared the optimal solutions from different methods in Fig. \ref{Penta-Results}a by evaluating the DBI score. Interestingly, our optimal KC clustering also corresponds to the lowest DBI score, indicating a better clustering using this measure as well.

As the pentapeptide system is relatively small, significant structural differences cannot be identified within the identified small microstates.
Correctly identifying these microstates is an open problem, with significant implications on the dynamics of the MSM \cite{ToOptimize}.
We expect more significant differences between states for larger networks that correspond to larger systems, going beyond prototype peptide simulations. 

In this example, our method finds the optimal clustering less frequently than in the synthetic examples of SBM networks. This could be due to the increased number of low $\Delta K$ partitionings available, which provide many local minima which are more difficult to escape.
However, this is not a problem unique to the GNN method, as the PTVC method encounters  the same issue, having difficulty finding the relevant areas of partitioning space and is unable to find the optimal clustering in this example using random initialization. 
The dense and interconnected nature of the nodes means it is difficult for PCCA+-based methods to identify the most optimal clustering as well, as more optimal clusters may not differ much in terms of meta-stability. 
The larger models suffer from a vanishing gradient due to the heavily conditional nature of optimizing all node partition probabilities at the same time, resulting in an inability to find the optimal clustering even once while nevertheless providing a better result on average than the GraphSAGE and LD model.
The best clusterings found by the GraphSAGE and TD models are shown in Fig. S4 for reference.
Still, they are outclassed by the PCCA+, and all but the GraphSAGE with linear decoder model provide worse performance than PTVC, highlighting the efficiency of simpler networks for this kind of task.

While further analysis of learning rate adjustments could provide an effective solution to train larger and more complex models on this problem, it is clear that training simple networks is both quicker and more reliable in finding the optimal clustering.
Feature comparisons on the GraphSAGE + LD example show the success of the method in finding the optimal partitioning over 1000 individual runs, with all features tested except for using the eigenvectors (Fig. \ref{Penta-Diagram}b).
MFPTs show the best performance here, finding the optimal solution the greatest number of times. Most likely due to the computational simplicity in finding effective mappings from the MFPTs to optimized partitionings, as the value of Eq. (\ref{loss_function}) directly depends upon them.
The poor performance of the eigenvectors is most likely down to numerical stability, as some of them are very small 64-bit floats.

One advantage of the PTVC method, which uses an MCMC-based algorithm, over our method, is the ability to be initialized from any arbitrary partitioning. 
The results shown are for the PTVC when given no initial clustering information, however, when the method is initialized from the PCCA+ solution it consistently finds the optimal solution on all occasions. 
In the GNN models, we also changed our initialization function $\mathcal{I}$ by pretraining the models to replicate the PCCA+ solution using mean-squared error loss that is minimal for the optimal PCCA+ solution. Once the GNN is trained to this specific loss function and identifies the PCCA+ clustering, it is then further trained using the Kemeny loss of Eq. (\ref{loss_function}). This pretraining, however, did not improve the performance of the GNN optimization (data not shown).

\begin{figure}[H]
	\centering
	\includegraphics[width=0.8\textwidth]{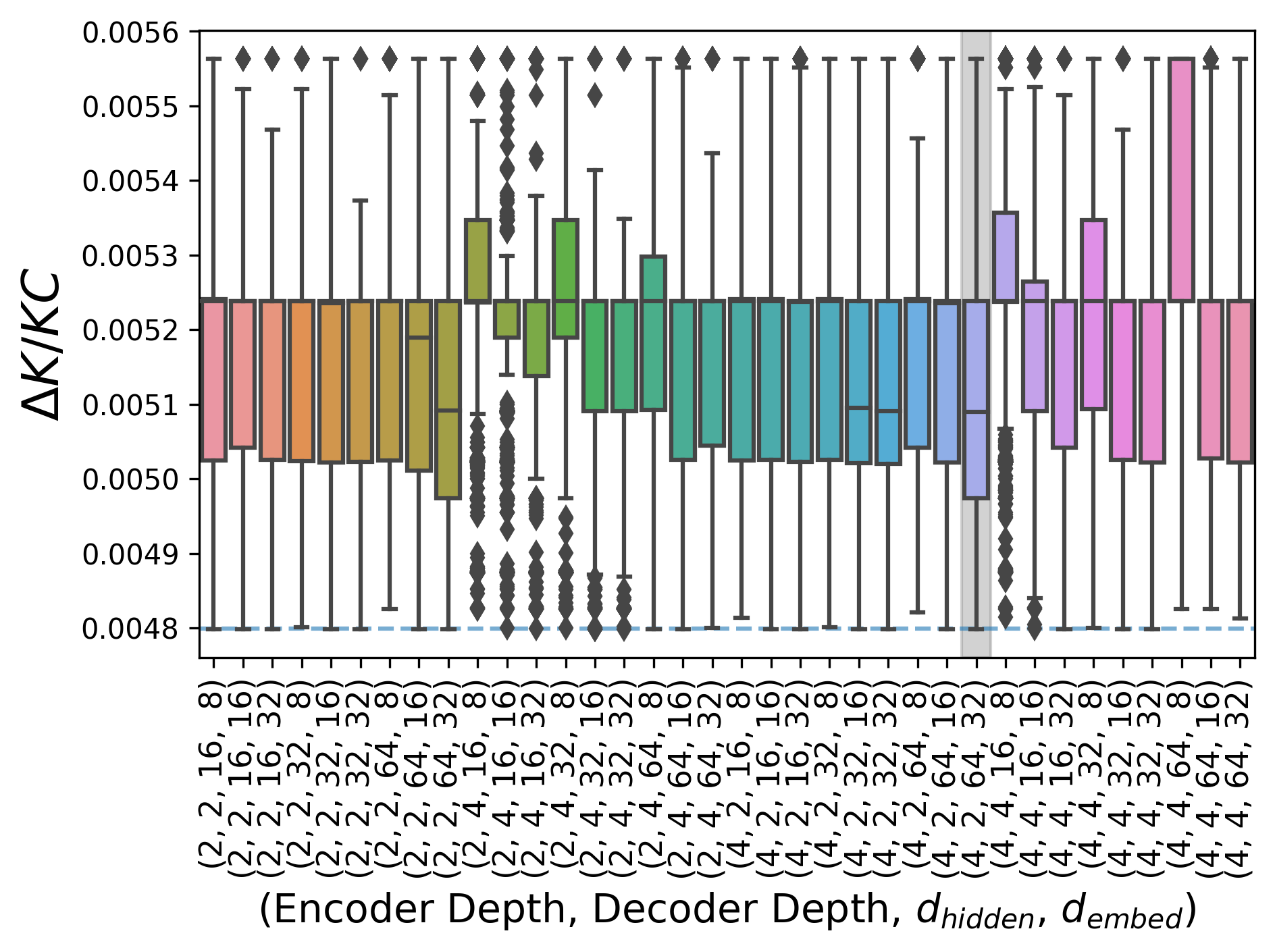}
	\caption{
		Parameter grid search on the pentapeptide example. 1000 networks were initialized for each set of parameters, and the resulting $\Delta K / KC$ distribution is shown. The grey background indicates the parameters identified as optimal.
	}
	\label{HPSearch}
\end{figure}

Fig. \ref{HPSearch} shows the results of a hyperparameter search using the best-performing simple GNN model to identify the optimal parameters for the network on the pentapeptide example. 
It highlights the problem that the network's effectiveness is highly dependent upon the chosen hyper-parameters of the neural network and the size of the network. The best parameter space for a given problem must be sufficiently large to be able to express the wide multitude of possible $\mathbf{S}$ matrices, while also being small enough to allow for effective optimization in a feasible time. This is highlighted by the poorer performance of most networks featuring four linear layers in the decoder, where most of the time the method either fails to find the optimal solution or does so rarely it is classified as an outlier. Future work can further verify whether a varying or learnable adjustment to the learning rate could provide a reasonable improvement, as this would allow networks to more efficiently probe the relevant regions of parameter space. Alternatively, further work could evaluate if it is possible to quantify the relationship between the size of the graph being clustered and the most optimal network size.

\begin{table}[H]
\caption{Network architecture comparison for the pentapeptide system. The four network architectures are compared in terms of the mean (standard deviation) and minimum for 1000 experiments, along with the percentage of times the optimal solution was found.}
\label{table:Penta}
\begin{tabular}{|l|c|c|c|}
\hline
\textbf{Model Name} & \multicolumn{1}{l|}{\textbf{Mean}} & \multicolumn{1}{l|}{\textbf{Minimum}} & \multicolumn{1}{l|}{\textbf{Global Minimum Accuracy}} \\ \hline
GraphSAGE + LD          & 271.03 ( $\pm$ 4.80)                                   & \cellcolor[HTML]{FAECCF}\textbf{258.67} & \cellcolor[HTML]{FAECCF}\textbf{2.8\%} \\
GATv2 + LD      & 273.28 ( $\pm$ 5.71)                                   & 262.81                                  & 0.0\%                                  \\
GraphSAGE + TD      & \cellcolor[HTML]{FAECCF}\textbf{266.66 ( $\pm$ 5.94)}  & 259.098                                 & 0.0\%                                  \\
GATv2 + TD & 274.92  ( $\pm$ 5.81)                                 & 262.74                                  & 0.0\%                                  \\ \hline
PCCA+ & 258.86  ( $\pm$ 0.0)                                 & 258.86                                & N/A                               \\ \hline
\end{tabular}
\end{table}

\newpara
\section{Conclusion}
We describe a GNN-based network clustering optimization method, which uses the computing infrastructure and tooling that has risen around GNNs to provide a working basis for difficult optimization problems on graphs. The benefits are numerous, as GNNs provide an ample Ansatz for beginning optimization for partitioning phenomena in situations where otherwise finding an efficient representation for the task proves difficult.

We provide a novel implementation of a GP approach to cluster graphs based on kinetic optimization criteria. Here we use the KC to obtain metastable kinetically optimal clusters. 
We outline a general methodology for training GNNs to optimize graph partitionings and compare the performance of four different architectures across three different kinetic network types, using several different node features.
Our results show that smaller and simpler networks often show superior performance in our test cases, as larger networks become more difficult to optimize, and often cannot find the optimal solutions. However, as we are only interested in the best solution with the maximal Kemeny constant, multiple evaluations can lead to the correct global solution, even if only obtained less frequently in individual initializations.
GraphSAGE layers combined with a linear decoder achieved both greater consistency and accuracy than any tested attention-based alternatives.

To show that our proposed method provides a viable alternative to existing methods in terms of the coarse grained systems, we compare optimal clusterings obtained with those identified using the PCCA+ clustering.
It is important to note, however, that as PCCA+ uses a different kinetics-based objective function, the task of identifying the optimal clustering using the KC-based measure is independent of this comparison. 
Testing the method on SBM graphs generated to have a clear optimal partitioning, we can attain similar results to PCCA+, but with more optimal values of the Kemeny constant. 
A linear chain corresponding to an analytic 1D free energy profile also poses a difficult task for GNNs due to the low connectivity, while PCCA+ in this case can produce the optimal solution.
We show how we can adapt the node features the GNN uses and the number of runs to alleviate this problem, and improve the performance.
However, the optimal clustering is only found in about 5\% of the attempts in only one GNN architecture, and therefore improving the reliability of the method still requires future work.
The outline of our general methodology, introducing the strategy defining generator $\mathcal{I}$, culling $\mathcal{C}$, and termination $\mathcal{T}$ functions, provides an ample framework for future work in this direction.

Our MD-simulation derived MSM for a pentapeptide is the most complex example we considered here. It has a slightly more optimal $\Delta K$ clustering that seems to better capture metastable states than the PCCA+-derived clusters, as also seen by comparing the corresponding DBI scores. Only one of the four architectures can find this optimal solution in less than 3\% of the time, highlighting the problem of GNNs, which further work on $\mathcal{I}$ functions can hopefully provide a solution for. Additionally, GNNs do not provide a straightforward way to initialize the search from a close-to-optimal solution, which makes the PTVC algorithm more efficient.

In this work, we explored a number of node features that were used by the GNNs to enable efficient optimization. The MFPT-based features were some of the most successful ones, as these directly relate to the definition of $\Delta K$. However, for examples derived from biomolecular simulations, features that are specifically relevant to physical and structural observables (angles, distances, energy terms, etc.) could outperform the general features used here. This is a promising avenue for future directions.

In summary, we have shown the validity of using GNNs as an effective method to solve the GP problem applied to MCs, alongside describing a general framework outlining the strategy for optimizing partitionings using GNNs in parallel on graph-like structures. 
We propose using such a framework to solve the many different and difficult non-convex optimization tasks which are commonplace in computational fields using graph structures, taking full advantage of the speed, availability, and efficiency of modern ML tooling and associated hardware.
The general training paradigm outlined provides a working basis for generalizing this strategy to different problems, and the results highlight the usefulness of simpler GNN architectures when solving such optimization problems.
It remains to be seen how this can be improved by implementing more complex parameter space sampling strategies to aid the identification of more optimal solutions, taking inspiration from similar techniques used for enhanced sampling in MD.

\section{Acknowledgments}

We thank Dr. Alessia Annibale for helpful suggestions, Madhav Sharma for his feedback on the manuscript, and Bin Liang for his initial contributions to pretraining. The authors acknowledge funding from ERC Starting Grant (Project 757850 BioNet) and EPSRC (grant no. EP/R013012/1). This project made use of time on ARCHER granted via the UK High-End Computing Consortium for Biomolecular Simulation, HECBioSim (http://hecbiosim.ac.uk).
The authors acknowledge the use of the UCL Myriad High Performance Computing Facility (Myriad@UCL), and associated support services, in the completion of this work.
\bibliography{all}
\end{document}

\maketitle
\thispagestyle{fancy}
% \section{Supporting Information}
\setcounter{figure}{0}
\renewcommand{\figurename}{Figure}
\renewcommand{\thefigure}{S\arabic{figure}}

\newpage
\section{Featurization}
In order learn effectively, GNNs require a distinct feature vector $F_i$ associated with each node $N$, which is used as the initial input to the model.
There are several different ways to produce $N$ feature vectors of dimension $d_{\text{feat}}$. Here, we aimed to use general features based on the properties of the MSM, and did not include system specific information, such as structural properties, which could also be used for molecular MCs. 
Below is a list of the different featurization methods we used for KC-based clustering.

\begin{itemize}

\item \textbf{Markov/Rate Matrix:} \\
As a MC can be fully defined by either a Markov matrix $\textbf{Q}$ or rate matrix $\mathbf{K}$, both of shape \\ $(N \times N)$, the simplest featurization is to use the columns of either defining matrix corresponding to the outgoing rates/transition probabilities as the $N$ feature vectors for each node with $d_{\text{feat}} = N$. 
This produces a set of features for each node $\mathbf{F_i}$ where each element of the feature matrix is simply a column of the Markov or rate matrix: $\mathbf{F_{i}} = (Q_{1i}, ...,Q_{Ni})$, or analogously, \\ $\mathbf{F_{i}} = (K_{1i}, ...,K_{Ni})$.  \\
We also tested whether performing PCA on the matrix $\mathbf{K}$ or $\mathbf{Q}$ prior to splitting into individual $F_i$ would improve performance, as done in other work on clustering using GNNs\cite{gap}.
To do so, we perform PCA using the Sci-Kit Learn Python library \cite{pedregosa2011scikit, 1}. 
Interpreting each column $\textbf{F}_i$ as a sample and the rows as features, we take all components of the PCA. 
This can be seen as a linear rotation of the defining matrix onto the perpendicular directions along which the node transition probabilities or rates vary the most, with the aim of making nodes with similar transitions appear more similar.

\item \textbf{Adjacency Matrix:} \\
As in the main work, we interpret the adjacency matrix of an MC to be a simple cutoff of the defining matrix $(Q_{ij}+Q_{ji})/2 > c = 0.015$. 
This yields a binary $(N \times N)$ square matrix, which we split into columns representing each node with $d_{\text{feat}} = N$. We also again tested a PCA of this, utilizing the same strategy as before.

\item \textbf{MFPT Matrix:} \\
As the MFPTs are one of the most relevant properties for the KC, we tested using them as node features. 
Firstly, the MFPT matrix with elements $t_{ji}$ of the unclustered graph was calculated using Eq. (12). This generates an $(N \times N)$ matrix of the MFPTs between nodes in the network.
If needed, we can then extract the PCA features, before finally splitting the matrix up into feature vectors $F_i$ of size $d_{\text{feat}} = N$ using the columns of the MFPT matrix.

\item \textbf{Eigenvectors:} \\
As the eigenvalues and eigenvectors of the Markov matrix describe important kinetic properties, we also used the eigenvectors of the Markov or rate matrices as feature vectors.

\item \textbf{Trainable Embedding:}
Trainable embeddings have found much success in Large Language Models (LLMs), where they provide an efficient way to represent the sentiments of different tokens, and have been employed previously as random features for GNNs.
Trainable embeddings are equivalent to an associative array data structure, where a predefined set of keys each point to a vector of arbitrary size. 
The vector is initialized with random numbers which are then trained along with the rest of the network during back-propagation steps.
We used the torch implementation\cite{torch} , which uses integer keys from $\{1 \dots N\}$ representing each node. Several embedding sizes were tested before settling on $d_{\text{feat}} = 128$ which provided the best performance across all systems.

\item \textbf{Single layer linear network:} \\
A single feedforward layer that takes as input for each node the $3N$-dimensional vector that is formed  by concatenating the corresponding columns of the (i) Markov/rate, (ii) adjacency, and (iii) MFPT matrices.
The output dimension was $d_{\text{feat}} = 128$ in all systems, and the weights of this layer were updated as part of the training process.
\end{itemize}

\newpage

\section{Additional Figures}
\begin{figure}[H]
\centering
\includegraphics[width=1\textwidth]{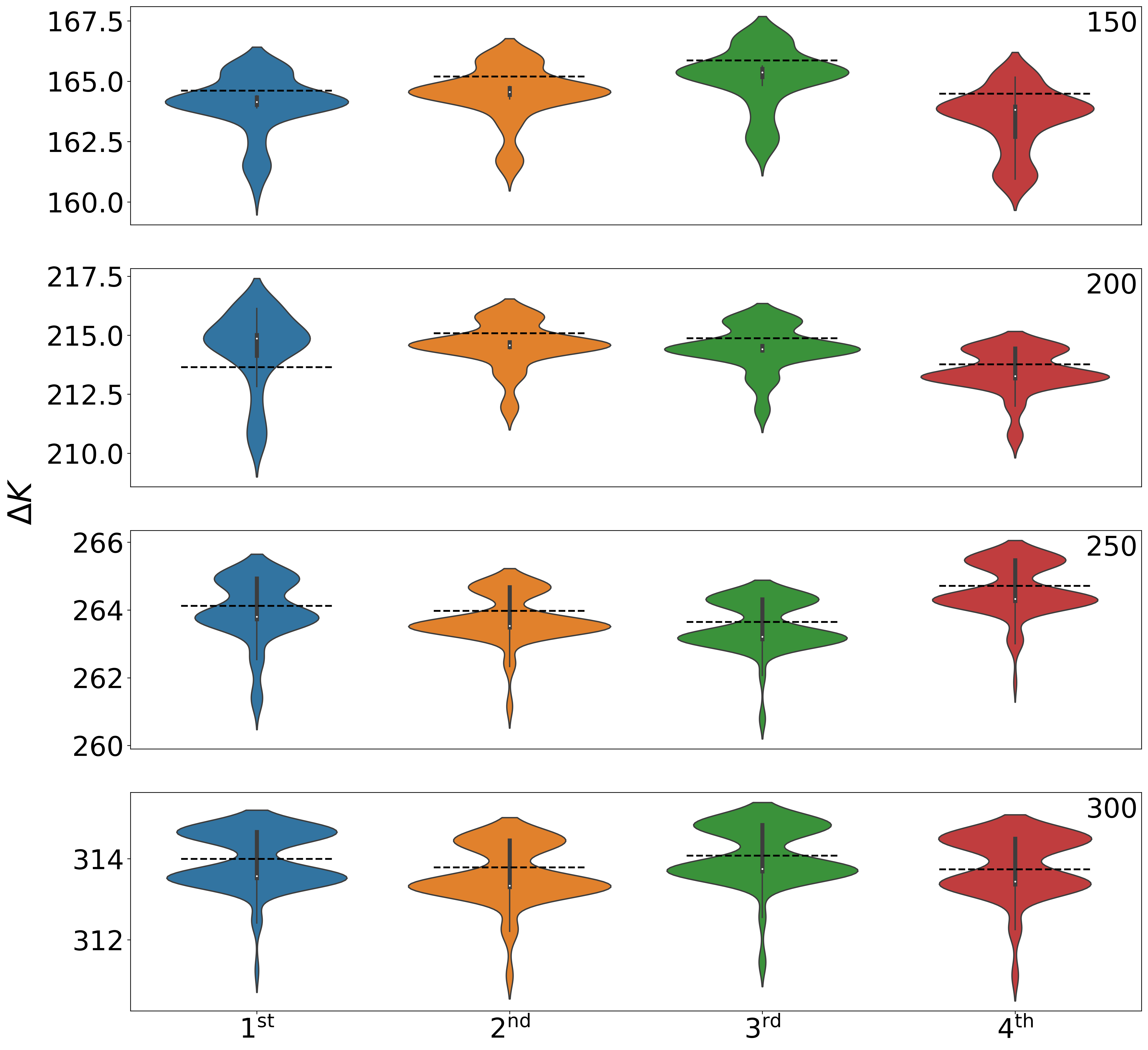}
% {\bf Figure S1.} 
\caption{A comparison of the $\Delta K$ values of the partitionings found by PCCA+ (the dotted black line) and the GraphSAGE + LD GNN method on four SBM networks, where the GNN provides a clustering with the most optimal value. These four networks represent the largest difference in our examples between the Kemeny optimum and the $\Delta K$ value corresponding to the PCCA+ clustering.}

\end{figure}

\begin{figure}[H]
	\centering
	\includegraphics[width=1\textwidth]{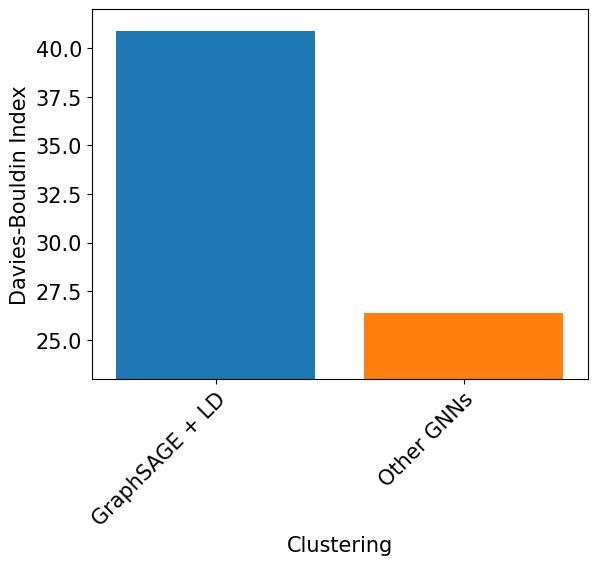}
	\caption{
		DBI values corresponding to clusterings found for the 1D example.
	}

\end{figure}

\begin{figure}[H]
	\centering
	\includegraphics[width=1\textwidth]{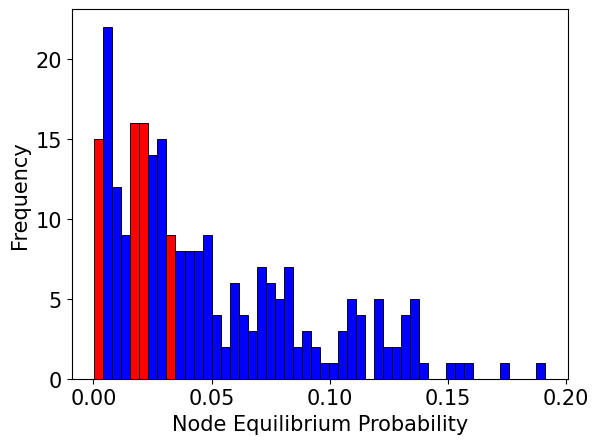}
	\caption{
		A histogram of the equilibrium probability distributions of each node in the pentapeptide example. Red bars indicate the presence of a node which is placed into a different cluster by PCCA+ and the GraphSAGE + LD method. These differing nodes represent 7.4\% of the overall node populations. 
	}

\end{figure}

\begin{figure}[H]
	\centering
	\includegraphics[width=1.3\textwidth]{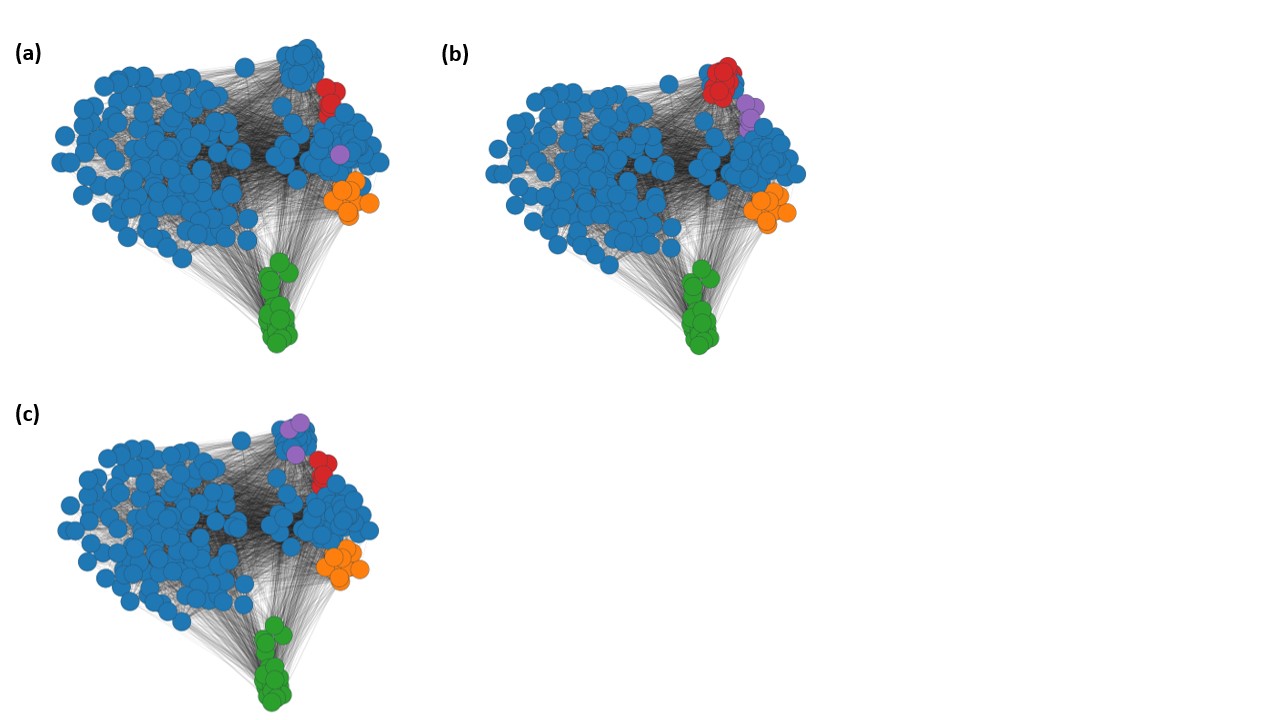}
	\caption{
		A comparison of the pentapeptide network partitionings with the lowest $\Delta K$ value found by each method used. Figures \textbf{(a)}, \textbf{(b)}, \textbf{(c)} show those found by the GATv2 + TD, GraphSAGE + TD, and GATv2 + LD respectively
	}

\end{figure}

\bibliography{si}